\documentclass[10pt,conference]{IEEEtran}
\IEEEoverridecommandlockouts
\usepackage{amsmath,amssymb,amsfonts}
\usepackage[numbers,sort&compress]{natbib}
\usepackage{algorithmic}
\usepackage{graphicx}
\usepackage{textcomp}
\usepackage{fancyhdr}
\usepackage{xcolor}
\usepackage{url}
\usepackage[colorlinks=false, urlcolor=blue, linkcolor=black]{hyperref}
\usepackage{subcaption}
\def\BibTeX{{\rm B\kern-.05em{\sc i\kern-.025em b}\kern-.08em
    T\kern-.1667em\lower.7ex\hbox{E}\kern-.125emX}}

\fancypagestyle{specialfooter}{%
  \fancyhf{}
  
  \fancyfoot[R]{ \noindent\fbox{%
    \parbox{\textwidth}{%
        {\footnotesize This work has been submitted to the IEEE for possible publication. Copyright may be transferred without notice, after which this version may no longer be accessible.}
        }
    }}
}
 
\begin{document}

\title{Quantum Patch-Based Autoencoder for Anomaly Segmentation}

\author{\IEEEauthorblockN{Maria Francisca Madeira$^*$}
\IEEEauthorblockA{\textit{Ludwig-Maximilians-Universität München} \\
\textit{Fraunhofer IKS}\\
Munich, Germany \\
francisca.madeira@lmu.de}
\and
\IEEEauthorblockN{Alessandro Poggiali$^*$}
\IEEEauthorblockA{\textit{University of Pisa} \\
\textit{Fraunhofer IKS}\\
Munich, Germany \\
alessandro.poggiali@phd.unipi.it}
\and
\IEEEauthorblockN{Jeanette Miriam Lorenz}
\IEEEauthorblockA{\textit{Fraunhofer IKS} \\
\textit{Ludwig-Maximilians-Universität München}\\
Munich, Germany \\
jeanette.miriam.lorenz@iks.fraunhofer.de}
}

\maketitle
\thispagestyle{plain}
\pagestyle{plain}
\thispagestyle{specialfooter} 
\def\thefootnote{*}\footnotetext{The authors contributed equally to the work.}\def\thefootnote{\arabic{footnote}}

\begin{abstract}
Quantum Machine Learning investigates the possibility of quantum computers enhancing Machine Learning algorithms. Anomaly segmentation is a fundamental task in various domains to identify irregularities at sample level and can be addressed with both supervised and unsupervised methods.
Autoencoders are commonly used in unsupervised tasks, where models are trained to reconstruct normal instances efficiently, allowing anomaly identification through high reconstruction errors. While quantum autoencoders have been proposed in the literature, their application to anomaly segmentation tasks remains unexplored. 
In this paper, we introduce a patch-based quantum autoencoder (QPB-AE) for image anomaly segmentation, with a number of parameters scaling logarithmically with patch size. QPB-AE reconstructs the quantum state of the embedded input patches, computing an anomaly map directly from measurement through a SWAP test without reconstructing the input image.
We evaluate its performance across multiple datasets and parameter configurations and compare it against a classical counterpart.
\end{abstract}


\section{Introduction}

Anomaly detection, one of the most crucial tasks in machine learning (ML), aims to identify irregularities or abnormalities within a dataset \cite{chandola_anomaly_2009}. In classical ML, anomaly detection can be solved using a supervised approach, where a classifier learns to distinguish between labeled normal and anomalous data. In real-case scenarios, datasets are often imbalanced due to the reduced number of anomalies over a huge distribution of normal data. Since anomalies often occur with low frequency and can even be initially unknown, supervised learning is not a preferred or feasible choice in this context. Unsupervised and semi-supervised approaches, where training data contain normal instances only, have been proposed in the literature \cite{goldstein2016comparative}. In this setting, models are trained to autonomously learn functions that know how normal data looks like \cite{pang_deep_2022}. Although accurate anomaly detection is essential in various domains, there is often a demand for precise information about the specific regions within the data that deviate from the norm. This task is referred to as anomaly localization or segmentation, where the goal is not only to signal the presence of an anomaly but also to localize it within the given data sample \cite{tao_deep_2022}. In this paper, we focus on image data, wherein anomaly segmentation involves obtaining a pixel-wise anomaly score so that it can also be seen as an anomaly detection task at the pixel level \cite{yang_visual_2022}.
This score can be obtained through several methods \cite{baur_autoencoders_2021}. Reconstruction based methods obtain the pixel-wise difference between the input and output of a reconstruction-based model, where it is expected that anomalous regions will produce local reconstruction error that surpasses a given threshold \cite{crimi_deep_2019, Chen2018UnsupervisedDO, shen_unsupervised_2019, chen_unsupervised_2020}.
\par Among the reconstruction-based models, a popular approach is the autoencoder \cite{hinton2006reducing}. This model demonstrates the capacity to effectively learn essential data features without requiring any supervision. Essentially, the autoencoder achieves self-replication (reconstruction) through an encoder-decoder network. The encoder component derives significant features from the input, while the decoder reconstructs the input based on these encoded features. The acquired representations are condensed within a bottleneck layer, referred to as the latent space, which usually has a much lower dimensionality than the input data. This ensures that only useful features are learned by the autoencoder, instead of merely copying the input
data for reconstructing the output \cite{tschannen2018recent}. 

It is expected that several machine learning tasks will benefit from the advancements of quantum computing in the future. Quantum machine learning (QML) emerges as an interdisciplinary domain, intersecting quantum computing (QC) and machine learning (ML)~\cite{biamonte2017quantum, schuld2015introduction}. One of the goals of QML is to leverage quantum properties to tackle classical ML tasks, investigating the possibility of quantum computers improving ML algorithms. The ability for a quantum computer to outperform a classical computer in a given task is referred to as \emph{quantum advantage}. While quantum advantage usually refers to a computational speedup, this term becomes broader in machine learning \cite{schuld_is_2022}, extending to improvements in terms of generalization, model expressivity, or quality of model predictions.


\par In supervised learning scenarios, research indicates that quantum machine learning models exhibit enhanced generalization compared to classical machine learning models when faced with limited training data \cite{caro2022generalization}. However, the extension of these findings to unsupervised learning remains an interesting open question. Although the notion of generalization does not directly translate to unsupervised learning tasks, attempts have been made to conceptualize it \cite{hansen_unsupervised_1996, appice_generalization_2015}. Segmentation tasks are particularly relevant in industrial and medical settings, where data is often scarce. Therefore, it is relevant to ask whether quantum computing could provide any practical advantages in this context.
\par In this work, we introduce a Quantum Patch-Based Autoencoder (QPB-AE) designed to address the anomaly segmentation task in images. Our method is tailored to meet the demands of Noisy Intermediate-Scale Quantum (NISQ) devices \cite{preskill_quantum_2018}, with encoding resources and parameter count growing logarithmically with the size of the patch. Furthermore, our trainable model operates entirely on the quantum level, enabling the direct acquisition of anomaly scores from measurement, making it feasible for implementation on real quantum hardware. Our method relies solely on classical computing for optimization and data post-processing tasks. Finally, unlike typical autoencoder methods, we do not explicitly reconstruct the input image. Instead, we leverage quantum properties in a resource-efficient manner to generate anomaly maps from the input with minimal classical post-processing.

We evaluate the performance of QPB-AE across multiple datasets. The results demonstrate that QPB-AE effectively localizes anomalies in images, returning satisfactory results in terms of the considered metrics. Moreover, comparing our quantum model with a classical counterpart, it turns out that QPB-AE exhibits superior performance in identifying anomalies despite its significantly lower number of trainable parameters and with a limited number of training images.

\par Our contributions are the following:
\begin{itemize}
\item We propose the first application of a quantum autoencoder-based method for the task of anomaly segmentation.
\item We propose a method that shows a logarithmic scaling of trainable parameters with the patch size, a hyperparameter that can be adjusted taking into account the input size and desired anomaly map resolution. Our proposed model is compatible with current quantum hardware, satisfying the constraints of Noisy Intermediate-Scale Quantum (NISQ) devices.
\item We benchmark our model on downsampled real-world medical and industrial datasets and we evaluate its performance with metrics according with those currently used in state of the art anomaly segmentation benchmarks.
\end{itemize}

The rest of the paper is organized as follows. Section \ref{sec:relatedwork} discusses related work. Section \ref{sec:model} introduces the model architecture. In Section \ref{sec:exp}, we evaluate the proposed model across multiple datasets and parameter configurations.
In Section \ref{sec:results}, we discuss the obtained results, and in Section \ref{sec:conclusion}, we derive conclusions.

\section{Related work}
\label{sec:relatedwork}

\par The first model of a quantum autoencoder, proposed by Romero et al. \cite{romero_quantum_2017}, aimed at efficient quantum data compression. 
This work laid the foundation for various research directions, with quantum autoencoders finding applications across a broad spectrum of tasks. These include anomaly detection on quantum data \cite{ngairangbam_anomaly_2022, kottmann_variational_2021}, quantum data denoising \cite{bondarenko_quantum_2020}, quantum error correction \cite{locher_quantum_2023} and efficient image compression \cite{wang2024quantum}.

Several works propose hybrid quantum-classical autoencoder architectures \cite{orduz2021quantum, srikumar2021clustering, sakhnenko2022hybrid}. In these approaches, quantum and classical trainable architectures are combined to leverage the strengths of both methods. 
Within this scope, a quantum convolutional autoencoder is proposed in \cite{orduz2021quantum}, where the model's reconstruction abilities and latent space representation are studied.
In \cite{srikumar2021clustering}, a hybrid quantum autoencoder is introduced to improve clustering and classification tasks. The authors in \cite{sakhnenko2022hybrid} propose a hybrid classical-quantum autoencoder specifically for anomaly detection in tabular data. 
\par While quantum autoencoders have shown promise in anomaly detection tasks, other works using different architectures have been proposed. In \cite{wang2023quantum}, the feasibility of quantum machine learning for deep image anomaly detection is explored. The authors propose a quantum-classical hybrid deep neural network that learned from normal raw images to train a normality model, detecting anomalies during inference. Anomaly detection in images extends to various domains, such as medical imaging and surveillance. For instance, a hybrid classical-quantum convolutional neural network classifier is introduced for stenosis detection in \cite{ovalle-magallanes_hybrid_2022}. In surveillance applications, quantum convolutional neural networks are applied for anomaly detection in videos \cite{amin_detection_2023}.

\par Quantum anomaly detection techniques are further extended to anomaly segmentation tasks, although this direction remains mostly unexplored. For medical images, \cite{amin_secure_2022} focuses on brain MRI image segmentation using a two-qubit quantum model. In \cite{pramanik_quantum-classical_2022}, classical and quantum processing blocks are applied to identify cracks on surface images. First, a quantum classifier is trained to recognize anomalous images, and then the images classified as anomalies are segmented using quantum clustering. Quantum image processing methods, such as those presented in \cite{caraiman_image_2015} and \cite{liu_quantum_2022}, offer unique approaches to segmenting and identifying anomalies within images using quantum principles.

The task of anomaly segmentation has, however, predominantly remained within the classical realm, with models proposing hybrid architectures relying solely on quantum circuits that function on a threshold-based approach and, thus, are not trainable and are unable to learn from data.
Furthermore, despite its success in anomaly detection tasks, no proposal to integrate an autoencoder-based anomaly segmentation model exists in the literature.

\section{Model}
\label{sec:model}
This section introduces the Quantum Patch-Based Autoencoder (QPB-AE) model for localizing image anomalies\footnote{Code available on Github: \url{https://github.com/AlessandroPoggiali/QPB-Autoencoder}}. 
The model receives an input image $X \in \mathbb{R}^\text{CxHxW}$, where C, H, and W represent the number of channels, height, and width of the input image, respectively, and returns an anomaly map $Y \in \mathbb{R}^\text{CxHxW}$.
\par The input image is processed patch by patch, each denoted as $x^p$ for $p=0,\dots,N-1$ (with $N$ the total number of patches). A patch $x^p$ corresponds to a square region of $N_P=P\times P$ pixels within the image $X$. Each patch is fed into a quantum encoder-decoder scheme, outputting a patch $z^p$ of the same size as $x^p$. The output patch has a uniform value, i.e, all pixels within the same patch $z^p$ are assigned a single \textit{similarity score}. The local similarity scores are then aggregated to form a similarity map $Z\in \mathbb{R}^\text{CxHxW}$. The processing involves defining a patch size $P$ and a stride $S$. We restrict ourselves to the case $H=W$ and $C=1$, for which the total number of patches is obtained as

\begin{equation}
N = \left(\left\lfloor\frac{H-P}{S}\right\rfloor + 1\right)^2.
\label{eqn:totalpatches}
\end{equation}
\par If the stride is smaller than the patch size, $S < P$, then the patches $z^p$ exhibit overlaps over $Z$. These are averaged over to obtain the final full map. Specifically, a pixel at a position $(i,j)$ on the similarity map $Z$, denoted $Z_{i,j}$ is obtained as:
\begin{equation}
Z_{i,j} = \frac{\sum^{N-1}_{p=0} \text{overlap}(i,j,p) \cdot z_{i,j}^p}{\text{count}(i,j)}
\label{eqn:fullmap}
\end{equation}
where $z^p_{i,j}$ is the similarity score of patch $p$ at position $(i,j)$, $\text{overlap}(i,j,p)$ is the number of times a patch $z^p$ overlaps with position $(i,j)$, and
$\text{count}(i,j)$ is the total number of overlaps over the position $(i,j)$. The final anomaly map can then be obtained as 
\begin{equation}
Y = 1 - Z.
\label{eqn:fullanomalymap}
\end{equation}
A smaller stride can be selected to obtain an anomaly map with a higher definition, at the cost of a larger number of circuit evaluations.

\subsection{Architecture}

\subsubsection{Image data embedding}


To encode classical patches into the quantum autoencoder, we take each patch $x^p$ to contain a number of pixels $N_P=2^n$ for a given $n$. Each patch $x^p$ is then flattened as a vector, normalized such that $\sum^{N_P-1}_{k=0} |x^p_k|^2 = 1$, and encoded into a patch quantum state of $n$ qubits, $|\phi\rangle$, as:
\begin{equation}
|\phi\rangle = \sum^{2^n-1}_{k=0} x^p_k |k\rangle
\label{eqn:quantumpatch}
\end{equation}
where $k$ denotes the $k$-th computational basis state.

\subsubsection{Encoder circuit and parameter scaling}
\par
\begin{figure}[htbp]
\centerline{\includegraphics[width=0.95\linewidth]{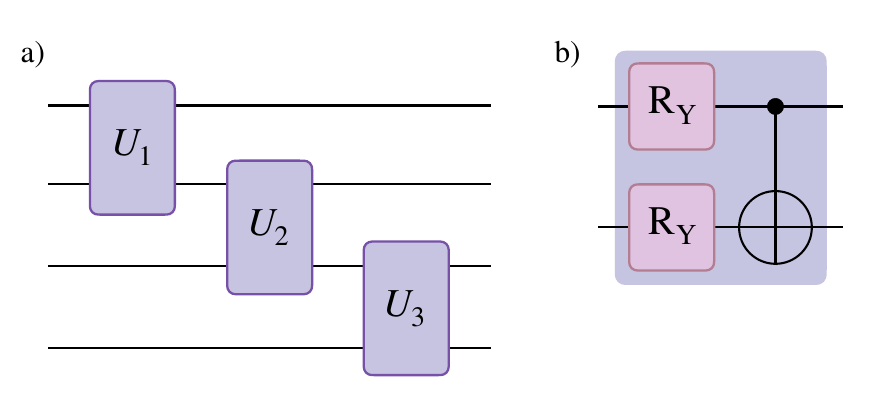}}
\caption{a) MPS ansatz, $U_{\text{MPS}}(\boldsymbol{\theta})$, used for compression of the patch quantum state. Each block on the MPS has the same structure, but different trainable parameters.
The circuit is used as the encoder on QPB-AE, while $U_{\text{MPS}}^{\dagger}(\boldsymbol{\theta})$ is used as the decoder.  
b) The chosen two-qubit MPS block, composed of two trainable $\text{R}_\text{Y}$ gates and a $\text{CNOT}$ gate. }
\label{fig:ansatzandblock}
\end{figure}

 The circuit we employ is based on \cite{romero_quantum_2017},
where quantum data is compressed through optimization of the fidelity between the original and reconstructed states.
We propose to use a tensor network Matrix Product State (MPS) \cite{perez-garcia_matrix_2007} ansatz as the encoder to compress the patch quantum state. This is motivated by the suggested efficiency of this architecture for quantum data compression \cite{dilip_data_2022} and by the observed high noise resilience of tensor network quantum circuits \cite{huggins_towards_2019}.
A schematic representation of the MPS circuit
$U_{\text{MPS}}(\boldsymbol{\theta})$ is shown in Fig. \ref{fig:ansatzandblock} a) for the case $n=4$.

\par 

Fig. \ref{fig:ansatzandblock} b) shows the chosen two-qubit unitary block, consisting of two trainable $\text{R}_\text{Y}$ gates and a $\text{CNOT}$ gate. Each block has 
$2$ trainable parameters, while
the number of blocks in the MPS ansatz is given by $\log_2{(N_P)}-1$.
The total number of trainable parameters is then given by
$n_{train} = 2(\log_2{(N_P)}-1)$,
scaling logarithmically $\mathcal{O}(\log_2{(N_P)})$ with the patch size $N_P$, and being independent of the image size. 

\begin{figure*}[t]
\centering
  \includegraphics[width=\textwidth]{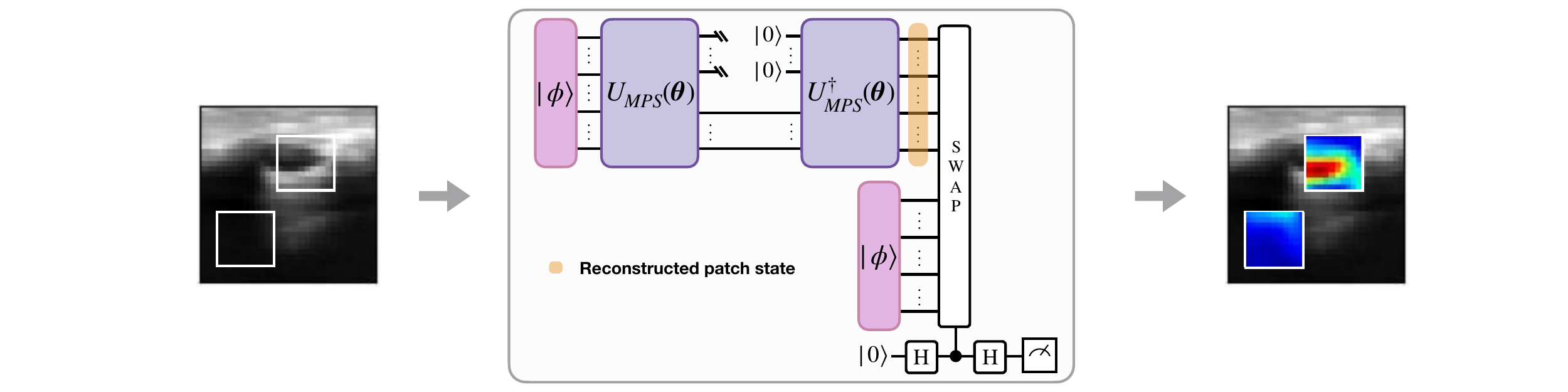}
  \caption{Schematic representation illustrating the full quantum autoencoder circuit (employed during the test phase to obtain the final anomaly maps). Each patch $x^p$ is encoded into a quantum state $|\phi\rangle$. The state is processed by the quantum autoencoder, that attempts to reconstruct the input wavefunction. A SWAP test between the input and reconstructed patch states yields a fidelity score that is post-processed to be interpreted as a probabilitic anomaly score.}
\label{fig:fullautoencoder}
\end{figure*}

\subsubsection{Full autoencoder architecture}
\par The full autoencoder circuit is shown in Fig. \ref{fig:fullautoencoder}. 
The goal of our model is to learn a compressed representation of the quantum patch state $|\phi \rangle$ down to a state of $n_{BD}$ qubits, where $BD$ denotes the \textit{bottleneck dimension} (an additional hyperparameter). The state is then decompressed to its original qubit count, resulting in a reconstructed quantum patch state. During training, the model is exposed to normal instances from the training set, aiming to accurately reconstruct the initial quantum state. In the presence of an anomaly, however, the model struggles to reconstruct the input wavefunction, leading to a decreased fidelity between the input and output states. This reduction in fidelity can be detected through measurement on a SWAP test, thus enabling local anomaly detection.

\par We now go over the details of the autoencoder construction. Consider two quantum systems $A$ and $B$, such that $|\phi\rangle_{AB}$ is the state of the composite system $AB$ on $n=n_{BD}+n_t$ qubits ($n_t$ denotes the number of \textit{trash} qubits). A successful compression of  $|\phi\rangle_{AB}$ by $U_{\text{MPS}}(\boldsymbol{\theta})$ implies a perfect disentanglement of $|\phi\rangle_{AB}$ into a compressed representation of $|\phi\rangle_{AB}$ in the subsystem $A$, which we denote 
$|c\phi \rangle_{A}$, and a trash state $|t \rangle_{B}$ of subsystem $B$:

\begin{equation}
U_{\text{MPS}}(\boldsymbol{\theta}) |\phi \rangle_{AB} = |c\phi \rangle_{A} \otimes |t \rangle_{B}.
\label{eqn:disentanglement}
\end{equation}

The trash state is mapped to a \textit{reference state} $|r \rangle_C$ on a system $C$, here $|r \rangle_C = |0 \rangle^{\otimes n_{t}}$. If such disentanglement occurs, then the state $|t \rangle_{B}$ can be efficiently discarded and we obtain a compressed representation of $| \phi \rangle_{AB}$ as a pure state, $|c\phi \rangle_{A}$. In this case, the action of $U^{\dagger}_{\text{MPS}}(\boldsymbol{\theta})$ will recover the original quantum patch state:

\begin{equation}
U^{\dagger}_{\text{MPS}}(\boldsymbol{\theta}) ( |c\phi \rangle_{A} \otimes |t \rangle_{B}) = U^{\dagger}_{\text{MPS}}(\boldsymbol{\theta}) U_{\text{MPS}}(\boldsymbol{\theta}) |\phi \rangle_{AB} = |\phi \rangle_{AB},
\label{eqn:decompression}
\end{equation}
and a SWAP test will yield a fidelity of 1. However, if such disentanglement does not occur, then the compressed state after discarding the subsystem $B$ is not a pure state. Instead, it can be described by the density matrix $\rho_A$, as
\begin{equation}
\rho_A = \mathrm{Tr}_B \left[ \left( {U^{AB}_{\text{MPS}}}(\boldsymbol{\theta}) |\phi \rangle_{AB} \right) \left( \langle \phi |_{AB}  {U^{AB}_{\text{MPS}}}^\dagger(\boldsymbol{\theta}) \right) \right].
\label{eqn:finalstate}
\end{equation}

\par The metrics we employ in this study take as input a probabilistic map where a high value corresponds to an anomaly and a low value corresponds to a normal instance (at pixel level). For this reason, we employ one last post-processing step (used only during the test phase), obtaining an estimate of the probability of measuring the state $|0\rangle$ from the measured expectation value $\langle \sigma_z 
\rangle$ on the SWAP test.

\subsection{Training and optimization}
\begin{figure}[h!]
\centerline{\includegraphics[width=0.99\linewidth]{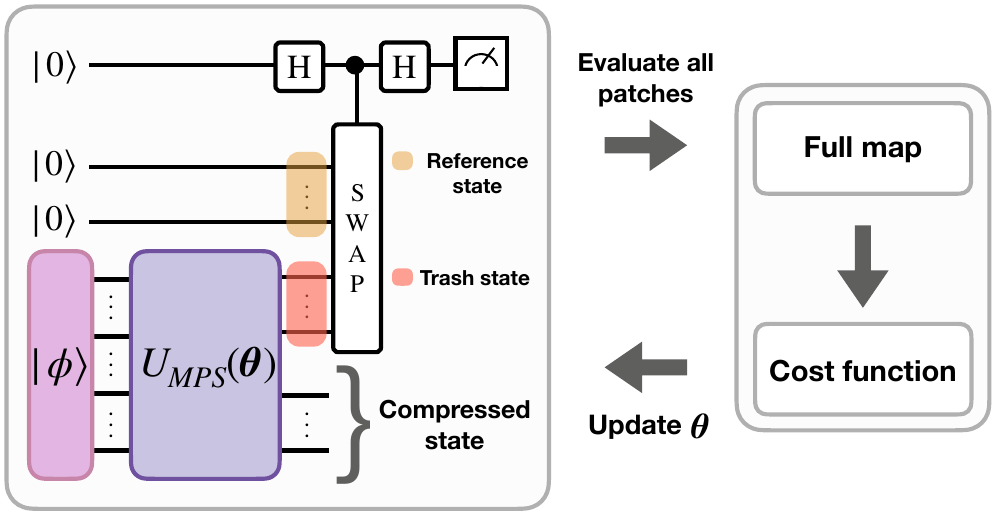}}
\caption{Schematic representation of the training process of QPB-AE. An optimal compression of the input patch $|\phi\rangle$ is obtained by maximizing the average fidelity between the trash state and the reference state over all patches. After all patches are evaluated, the full map is constructed, the cost function is evaluated and the parameters are updated.}
\label{fig:training}
\end{figure}

\par
As discussed above, the autoencoder tries to find a perfect disentanglement of $|\phi \rangle_{AB}$ in the subspaces $A$ and $B$ for normal instances. This is only possible if the trash state $|t \rangle_B$ is efficiently mapped to the reference state $|r \rangle_C = |0 \rangle^{\otimes n_{t}}$. It is therefore enough to optimize the encoder circuit by maximizing the fidelity between reference and trash states, so that the action of $U_{\text{MPS}}(\boldsymbol{\theta})$ on normal patch instances is as close as possible to Eq. \ref{eqn:disentanglement}
\cite{romero_quantum_2017}.
\par We therefore train the model using only the encoder part of the circuit, as exemplified schematically in Fig. \ref{fig:training}. After a given quantum patch state $|\phi\rangle$ is compressed, a SWAP test is performed between the reference state and the trash state. A measurement is performed to obtain the expectation value $\langle \sigma_z 
\rangle$, which is proportional to the fidelity between the trash and reference states. This value is thus interpreted as a similarity score and is assigned to the corresponding output patch $z^p$. This process is repeated for all patches and a full map $Z'$ is constructed according to Eq. \ref{eqn:fullmap}. Finally, we optimize a cost function that maximizes the average fidelity over all patches (taking into account possible overlaps), as
\begin{equation}
C(\boldsymbol{\theta}) = 1 - \frac{1}{H^2} \sum_{i=1}^{H} \sum_{j=1}^{W}  Z^{'}_{i,j}(\boldsymbol{\theta}).
\label{eqn:costfunction_modified}
\end{equation}

\par While the model processes the image sequentially, one patch at a time, the optimization process considers the entire image. The cost function is computed only after all patches have been evaluated and the full anomaly map has been constructed, enabling the model to learn local patch features with global image information. 

\section{Experiments}
\label{sec:exp}
We evaluate the performance of QPB-AE across multiple datasets and parameter configurations (54 configurations per dataset). The experiments are carried out using the Pytorch interface of Pennylane \cite{bergholm2018pennylane}, without considering noise. All experiments are performed training the model for 20 epochs using the Adam optimizer with a learning rate of 0.005 and a batch size of 4. The final results for each parameter configuration are averaged over three different random seeds. 
\subsection{Data}
For benchmark evaluation, we consider the MvTEC-AD \cite{bergmann2019mvtec} dataset, one of the most widely used anomaly detection and segmentation benchmarks. It consists of 15 categories of visual inspections, providing both normal and anomalous images, along with masks for the anomalies. In particular, we consider the categories Carpet, Leather, and Wood. Moreover, we test QPB-AE on the Breast Cancer images dataset \cite{al2020dataset} (BUSI), which consists of breast ultrasound images categorized into normal, benign, or malignant. It also provides the mask of the anomaly associated with every image. We merge the benign and malignant categories into a single anomalous class, comprising images with either benign or malignant cancers.
\par All datasets are used in a semi-supervised setting, training the model on normal images and using anomalies only for the test phase. 
We consider 100 images for training and 25 for validation for every dataset. 
As our test set, we use all the test images available for the MvTEC-AD datasets, whereas for BUSI, we evaluate the model over a subset of 100 test images, keeping the same benign/malign ratio as the original test set. Additionally, we compress the images down to a resolution of 32x32 pixels.

\subsection{Metrics}
Since QPB-AE outputs an anomaly map where high pixel values correspond to a high likelihood of containing anomalies, it is necessary to establish a threshold value, so that pixels exceeding this threshold within the map are classified as anomalies. Using the established thresholds, we employ the following metrics to assess the results:
\begin{itemize}
    \item The Intersection over Union (IoU) score, defined as 
    \begin{equation}
        \text{IoU(A,B)} = \frac{A \cap B}{A \cup B},
    \end{equation}
    \item The Dice score, defined as \begin{equation}
\text{dice(A,B)} = \frac{2 \times |A \cap B|}{|A| + |B|}.
\label{eqn:dice}
\end{equation}
\end{itemize}
Both metrics relate to a pixel-wise comparison between the identified anomaly and the ground truth, i.e., the associated mask. In particular, $A$ is the set of pixels corresponding to the identified anomaly and $B$ is the ground truth of the anomaly.  
\par We consider two threshold-independent metrics: the pixel-level Area Under the Receiver Operating Characteristic Curve (AUROC) \cite{fawcett2006introduction}, and the pixel-level Area Under the Per-Region-Overlap curve (AUPRO)\footnote{To compute the AUPRO, we adapt the implementation provided within Anomalib \cite{akcay2022anomalib}.} \cite{bergmann2019mvtec}. The metrics consider as the true positive rate (TPR) the percentage of pixels correctly classified as anomalies, and as the false positive rate (FPR) the percentage of pixels wrongly classified as anomalies for a given input image. The AUROC is then obtained by computing the area under the FPR-TPR curve, where several $\text{(FPR, TPR)}$ pairs are obtained 
by varying the classification threshold. 
\par Unlike the AUROC, which tends to favour large anomaly regions within images because a single accurately predicted large anomaly region could compensate for numerous falsely indicated small abnormal areas, the AUPRO assigns equal importance to varying-sized anomalies. This is done by computing the area under a PRO-FPR curve, where the PRO (Per-Region Overlap) is obtained for several values of the FPR in the range $[0, 0.3]$. The PRO is obtained by segmenting the ground truth mask into \textit{connected components}, i.e., different anomalous regions within the same instance. The PRO is then obtained as:

\begin{equation}
\text{PRO(A,B)} = \frac{1}{N} \sum_{i} \sum_{k} \frac{|A_{i} \cap B_{i,k}|}{|B_{i,k}|}
\label{eqn:pro}
\end{equation}
where $A_{i}$ defines the pixel-wise region predicted as anomalous, for a given threshold according to the target FPR value and for an image $i$, and $B_{i,k}$ denotes the anomalous ground truth pixel-wise region for the connected component $k$ in the image $i$.


\subsection{Parameters setup}
We analyze the results of QPB-AE over different parameters. In particular, we consider three different patch sizes $P=\{2,4,8\}$, and for each patch size different strides: ($S=\{1,2\}$ for $P=2$, $S=\{1,2,4\}$ for $P=4$, and $S=\{1,2,4,8\}$ for $P=8$). Moreover, for $P=4$ and $P=8$, we test two compression levels given by the bottleneck dimension, $BD$ (for $P=2$, $BD$ is kept fixed and is equal to $1$). Each choice of the bottleneck dimension leads to a different level of information compression. We quantify it by considering the compression percentage between the initial number of patch pixels $N_P=P \times P$, and the dimension of the statevector of the compressed patch, i.e., $N_P \rightarrow 2^{n_{BD}} (\%)$. We compare results using $BD=1$ (corresponding to a compression level of $87.5\%$ for $P=4$, and $96.87\%$ for $P=8$), and $BD=2$ (corresponding to a compression level of $75\%$ for $P=4$, and $93.75\%$ for $P=8$). 

\begin{figure*}[h!]
\centering
\includegraphics[width=0.65\textwidth]{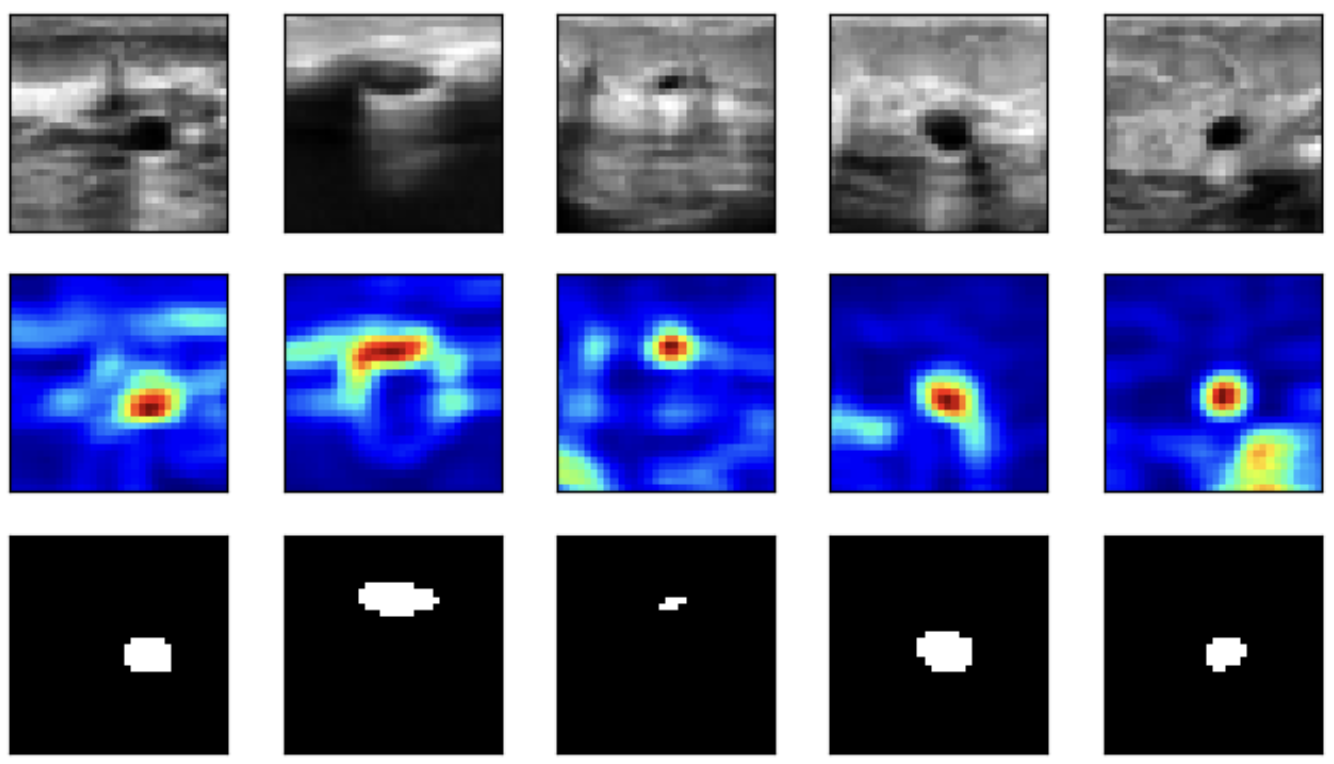}
  \caption{QPB-AE results on several breast ultrasound images (BUSI dataset), for $P=4$ and $S=1$. The top, middle, and bottom rows show the input images, predicted anomaly maps, and ground-truth masks, respectively.}
\label{fig:anomalymaps}
\end{figure*}

\subsection{Classical comparison}
\label{sec:classical}
We compare our quantum model with a classical counterpart architecture. For this, we focus on the case where $P=8$ and consider a basic, fully connected architecture following the autoencoder structure. To keep the same compression level as the QPB-AE with $P=8$ and $BD=2$ (93.75\%), the classical architecture has one input layer of 64 units for encoding the 64 pixels of the input patch $x^p$, one bottleneck layer of 4 hidden units, and one output layer of 64 units which 
returns a reconstruction $x'^p$ of the input patch. Each dense layer is followed by the ReLU activation function. A map is constructed based on the cosine similarity between input and output patches, defined as:

\begin{equation}
S_C(x^p,x'^p) = \frac{x^p \cdot x'^p}{||x^p||\,||x'^p||}
\label{eqn:cosine}
\end{equation}
This measure provides a similarity score consistent with the fidelity used by the QPB-AE. The cosine similarity score is therefore assigned to each output patch $z^p$. 
The anomaly map is constructed following Eqs. \ref{eqn:fullmap} and \ref{eqn:fullanomalymap}, and the loss function is given by Eq. \ref{eqn:costfunction_modified}. The model undergoes training using identical parameters for learning rate, optimizer, and epochs, and we consider the BUSI dataset for training and testing. The number of trainable parameters is in this case 132, while QPB-AE with the same hyperparameter configuration has only 10 trainable parameters.

\section{Results}
\label{sec:results}
As a demonstration of the QPB-AE output, Figure \ref{fig:anomalymaps} displays the anomaly segmentation outcomes for several BUSI images. The heatmaps highlight areas with a higher likelihood of anomalies, represented by red regions. 

In Table \ref{tab1}, \ref{tab2}, and \ref{tab3}, we report the results in terms of AUPRO and AUROC, for $P=2$, $P=4$, and $P=8$, respectively. The tables show that the best results are generally achieved when the stride is set to 1, except for $P=8$, where comparable results are also observed with $S=2$ and $S=4$. This aligns with the fact that a smaller stride leads to the higher resolution of the anomaly map at the cost of a greater number of patches to be fed into the model, thus resulting in higher execution costs. Regarding different compression levels specified by the $BD$ parameter, we see from tables that for $P=4$, we get better results when $BD=1$ (compression of 87.5\%), except for the Wood dataset. Conversely, for $P=8$, a compression of 93.75\% with $BD=2$ outperforms higher compression of 96.87\% for all datasets. 

\begin{figure*}
    \centering
    \begin{subfigure}{.5\textwidth}
      \centering
      \includegraphics[width=0.8\linewidth]{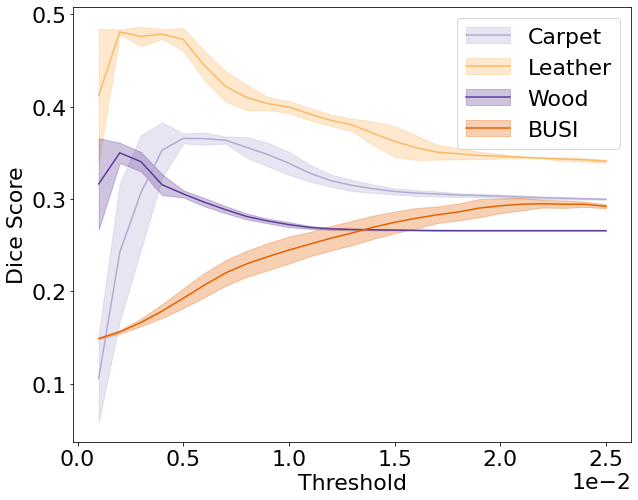}
      \caption{}
      \label{fig:sub1}
    \end{subfigure}%
    \begin{subfigure}{.5\textwidth}
      \centering
      \includegraphics[width=0.8\linewidth]{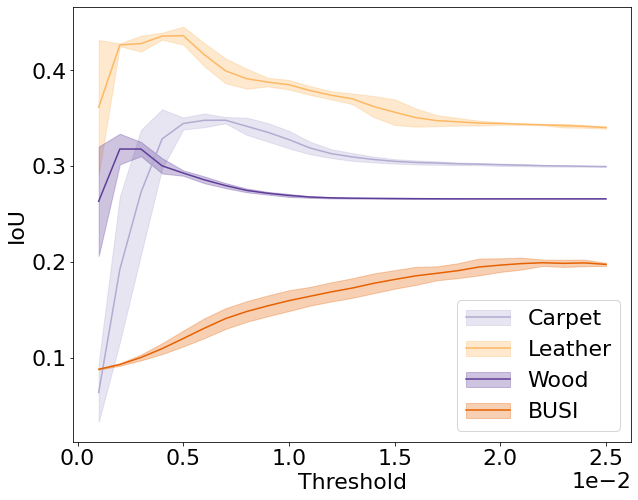}
      \caption{}
      \label{fig:sub2}
    \end{subfigure}
    \caption{Dice score (a) and IoU (b) for several threshold values across the evaluated datasets, for $P=4$, $S=1$ and $BD=2$.    
    }
    \label{fig:thresholds}
\end{figure*}

For the configuration $P=4$, $S=1$, and $BD=2$, which seems to perform the best on average in terms of both AUPRO and AUROC, we show in Figure \ref{fig:thresholds} how Dice score and IoU vary for different threshold values. In particular, since anomaly maps are strongly data-dependent, we can see that the peak values are reached at different thresholds for different datasets. This is more evident for the BUSI dataset, which has the maximum value for Dice and IoU at a much higher threshold compared to the other datasets. Furthermore, it can be observed that above a certain threshold, the value of both metrics for the MvTEC-AD datasets becomes constant. This is because the threshold is too high to discriminate between anomalous and normal pixels, and therefore every pixel is considered normal. The value in these cases is not zero because there are images without anomalies in the  MvTEC-AD test sets that have corresponding masks all zero, and thus in this case, both Dice and IoU are counted as 1. Best results in both Dice score and IoU are achieved for the Leather dataset, according to what is indicated by AUPRO and AUROC.

\begin{figure*}
    \centering
    \begin{subfigure}{.5\textwidth}
      \centering
      \includegraphics[width=0.8\linewidth]{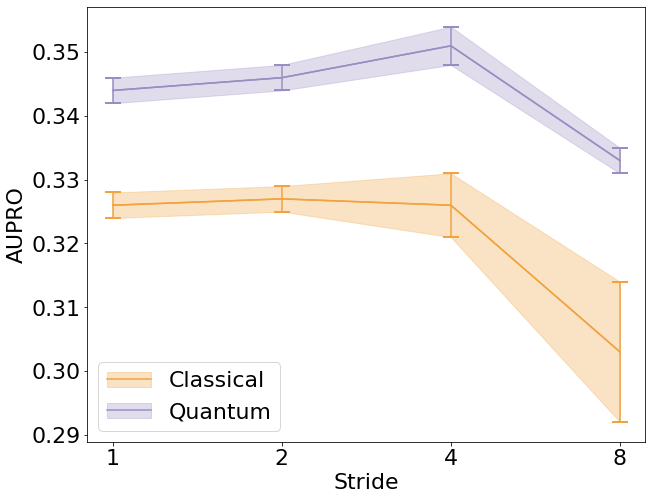}
      \caption{}
      \label{fig:sub1}
    \end{subfigure}%
    \begin{subfigure}{.5\textwidth}
      \centering
      \includegraphics[width=0.8\linewidth]{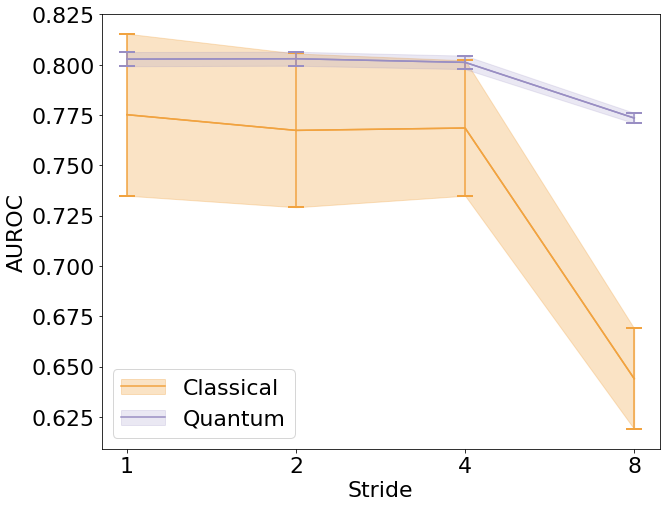}
      \caption{}
      \label{fig:sub2}
    \end{subfigure}
    \caption{Performance of QPB-AE against its classical counterpart in terms of AUPRO (a) and AUROC (b) for the case $P=8$ and a for compression level of $93.75\%$.}
    \label{fig:classical-vs-quantum}
\end{figure*}

Finally, Figure \ref{fig:classical-vs-quantum} compares the results between QPB-AE and the classical model described in Section \ref{sec:classical} on the BUSI dataset with $P=8$. For the particular configuration under consideration, QPB-AE demonstrates superior average performance for each stride despite having a significantly smaller number of trainable parameters. Moreover, the high number of parameters for the classical model leads to significant variance in performance across different weight initialization. In contrast, the results from QPB-AE are more consistent and stable.

\begin{table*}
\scriptsize
\caption{P = 2}
\begin{center}
\begin{tabular}{|l|c|c|c|c|}
\cline{2-5}
\multicolumn{1}{c|}{}&\multicolumn{2}{|c|}{\textbf{S = 1}} &\multicolumn{2}{|c|}{\textbf{S = 2}}\\ \cline{2-5}
\multicolumn{1}{c|}{}& \textbf{AUPRO} & \textbf{AUROC} & \textbf{AUPRO} & \textbf{AUROC} \\
\hline
Carpet & $0.492 \pm 0.140$ & $0.723 \pm 0.109$ & $0.397 \pm 0.120$ & $0.639 \pm 0.099$ \\
Leather & $0.867 \pm 0.131$ & $0.856 \pm 0.174$ & $0.756 \pm 0.168$ & $0.776 \pm 0.207$ \\
Wood & $0.558 \pm 0.147 $ & $0.677 \pm 0.141$ & $0.471 \pm 0.114$ & $0.620 \pm 0.123$ \\ \hline
BUSI & $0.545 \pm 0.049$ & $0.673 \pm 0.058$ & $0.465 \pm 0.049$ & $0.632 \pm 0.048$ \\
\hline
\end{tabular}
\label{tab1}
\end{center}
\end{table*}

\begin{table*}
\scriptsize
\caption{P = 4}
\begin{center}
\begin{tabular}{|l|c|c|c|c|c|c|}
\multicolumn{1}{c}{}&\multicolumn{6}{c}{\textbf{BD = 1}} \\ \cline{2-7} 
\multicolumn{1}{c|}{}&\multicolumn{2}{|c|}{\textbf{S = 1}} &\multicolumn{2}{|c|}{\textbf{S = 2}}&\multicolumn{2}{|c|}{\textbf{S = 4}}\\ \cline{2-7}
\multicolumn{1}{c|}{}& \textbf{AUPRO} & \textbf{AUROC} & \textbf{AUPRO} & \textbf{AUROC} & \textbf{AUPRO} & \textbf{AUROC} \\
\hline
Carpet & $\mathbf{0.729 \pm 0.046}$ & $\mathbf{0.908 \pm 0.014}$ & $0.697 \pm 0.042$ & $0.886 \pm 0.016$ & $0.571 \pm 0.041$ & $0.786 \pm 0.019$ \\
Leather & $\mathbf{0.953 \pm 0.002}$ & $\mathbf{0.990 \pm 0.001}$ & $0.951 \pm 0.004$ & $0.990 \pm 0.002$ & $0.925 \pm 0.026$ & $0.978 \pm 0.008$ \\
Wood & $0.389 \pm 0.008$ & $0.751 \pm 0.025$ & $0.382 \pm 0.007$ & $0.743 \pm 0.027$ & $0.368 \pm 0.003$ & $0.712 \pm 0.031$ \\ \hline
BUSI & $\mathbf{0.581 \pm 0.003}$ & $\mathbf{0.824 \pm 0.006}$ & $0.571 \pm 0.004$ & $0.820 \pm 0.005$ & $0.534 \pm 0.005$ & $0.794 \pm 0.005$ \\
\hline
\multicolumn{7}{c}{}\\
\multicolumn{1}{c}{}&\multicolumn{6}{c}{\textbf{BD = 2}} \\ \cline{2-7} 
\multicolumn{1}{c|}{}&\multicolumn{2}{|c|}{\textbf{S = 1}} &\multicolumn{2}{|c|}{\textbf{S = 2}}&\multicolumn{2}{|c|}{\textbf{S = 4}}\\ \cline{2-7}
\multicolumn{1}{c|}{}& \textbf{AUPRO} & \textbf{AUROC} & \textbf{AUPRO} & \textbf{AUROC} & \textbf{AUPRO} & \textbf{AUROC} \\ \hline
Carpet & $0.722 \pm 0.061$ & $0.906 \pm 0.002$ & $0.692 \pm 0.060$ & $0.885 \pm 0.026$ & $0.569 \pm 0.058$ & $0.790 \pm 0.029$ \\
Leather & $0.945 \pm 0.006$ & $0.987 \pm 0.003$ & $0.943 \pm 0.009$ & $0.986 \pm 0.004$ & $0.919 \pm 0.033$ & $0.973 \pm 0.014$ \\
Wood & $\mathbf{0.567 \pm 0.042}$ & $\mathbf{0.822 \pm 0.061}$ & $0.556 \pm 0.045$ & $0.814 \pm 0.063$ & $0.509 \pm 0.052$ & $0.774 \pm 0.069$ \\ \hline
BUSI & $0.584 \pm 0.005$ & $0.775 \pm 0.012$ & $0.570 \pm 0.008$ & $0.771 \pm 0.012$ & $0.516 \pm 0.012$ & $0.741 \pm 0.011$ \\
\hline
\end{tabular}
\label{tab2}
\end{center}
\end{table*}

\begin{table*}
\center
\scriptsize
\caption{P = 8}
\begin{center}
\begin{tabular}{|l|c|c|c|c|c|c|c|c|}
\multicolumn{1}{c}{}&\multicolumn{8}{c}{\textbf{BD = 1}} \\ \cline{2-9} 
\multicolumn{1}{c|}{}&\multicolumn{2}{|c|}{\textbf{S = 1}} &\multicolumn{2}{|c|}{\textbf{S = 2}}&\multicolumn{2}{|c|}{\textbf{S = 4}}&\multicolumn{2}{|c|}{\textbf{S = 8}}\\ \cline{2-9}
\multicolumn{1}{c|}{}& \textbf{AUPRO} & \textbf{AUROC} & \textbf{AUPRO} & \textbf{AUROC} & \textbf{AUPRO} & \textbf{AUROC} & \textbf{AUPRO} & \textbf{AUROC} \\
\hline
Carpet & $0.591 \pm 0.039$ & $0.878 \pm 0.019$ & $0.569 \pm 0.036$ & $0.867 \pm 0.019$ & $0.538 \pm 0.030$ & $0.846 \pm 0.018$ & $0.468 \pm 0.030$ & $0.779 \pm 0.016$ \\
Leather & $0.774 \pm 0.013$ & $0.966 \pm 0.003$ & $0.779 \pm 0.013$ & $0.967 \pm 0.003$ & $0.785 \pm 0.014$ & $0.967 \pm 0.004$ & $0.766 \pm 0.022$ & $0.951 \pm 0.008$ \\
Wood & $0.270 \pm 0.009$ & $0.749 \pm 0.009$ & $0.275 \pm 0.006$ & $0.748 \pm 0.007$ & $0.271 \pm 0.004$ & $0.738 \pm 0.008$ & $0.267 \pm 0.008$ & $0.704 \pm 0.013$ \\ \hline
BUSI & $0.339 \pm 0.003$ & $0.804 \pm 0.002$ & $0.341 \pm 0.003$ & $0.804 \pm 0.002$ & $0.348 \pm 0.003$ & $0.803 \pm 0.002$ & $0.334 \pm 0.002$ & $0.774 \pm 0.002$ \\ 
\hline
\multicolumn{9}{c}{}\\
\multicolumn{1}{c}{}&\multicolumn{8}{c}{\textbf{BD = 2}} \\ \cline{2-9} 
\multicolumn{1}{c|}{}&\multicolumn{2}{|c|}{\textbf{S = 1}} &\multicolumn{2}{|c|}{\textbf{S = 2}}&\multicolumn{2}{|c|}{\textbf{S = 4}}&\multicolumn{2}{|c|}{\textbf{S = 8}}\\ \cline{2-9}
\multicolumn{1}{c|}{}& \textbf{AUPRO} & \textbf{AUROC} & \textbf{AUPRO} & \textbf{AUROC} & \textbf{AUPRO} & \textbf{AUROC} & \textbf{AUPRO} & \textbf{AUROC} \\
\hline
Carpet & $0.616 \pm 0.043$ & $0.889 \pm 0.019$ & $0.597 \pm 0.042$ & $0.880 \pm 0.019$ & $0.557 \pm 0.036$ & $0.854 \pm 0.019$ & $0.484 \pm 0.041$ & $0.777 \pm 0.016$ \\
Leather & $0.779 \pm 0.013$ & $0.967 \pm 0.003$ & $0.783 \pm 0.013$ & $0.967 \pm 0.003$ & $0.789 \pm 0.013$ & $0.967 \pm 0.004$ & $0.775 \pm 0.022$ & $0.953 \pm 0.008$ \\
Wood & $0.297 \pm 0.002$ & $0.764 \pm 0.006$ & $0.303 \pm 0.003$ & $0.762 \pm 0.006$ & $0.302 \pm 0.003$ & $0.753 \pm 0.007$ & $0.281 \pm 0.003$ & $0.710 \pm 0.011$ \\ \hline
BUSI & $0.344 \pm 0.002$ & $0.803 \pm 0.003$ & $0.346 \pm 0.002$ & $0.803 \pm 0.003$ & $0.351 \pm 0.003$ & $0.801 \pm 0.003$ & $0.333 \pm 0.002$ & $0.773 \pm 0.003$ \\ 
\hline
\end{tabular}
\label{tab3}
\end{center}
\end{table*}

\section{Conclusion}
\label{sec:conclusion}

This paper introduced QPB-AE, a quantum autoencoder model for solving the anomaly localization task in images. 
Thanks to the efficient use of quantum resources and the low number of trainable parameters, the model is suitable for implementation in real NISQ architectures. Moreover, the experimental evaluation showed satisfactory results within the selected dataset. In particular, it is worth noting that the model achieved satisfactory results even with a limited number of training samples. As a potential future work, we aim to investigate whether the ability of an unsupervised quantum model to perform well with limited data is akin to the concept of generalization we have in a supervised scenario.

Finally, it is important to consider that since input images are initially down-scaled in resolution, the model worked with compressed data. This could impact result quality as anomaly localization becomes more challenging. We would expect an improvement in the quality of the results when working with higher-resolution images.

\section{Acknowledgments}


M. F. Madeira acknowledges funding by the Bavarian State Ministry of Science and the Arts with funds from the Hightech Agenda Bayern Plus, as a part of the funding initiative Munich Quantum Valley. A. Poggiali is member of the Gruppo Nazionale Calcolo Scientifico-Istituto Nazionale di Alta Matematica (GNCS-INdAM)  which provided a partial support for this work.

\bibliographystyle{IEEEtran}
\bibliography{biblio}

\end{document}